\documentclass[conference]{IEEEtran}
\IEEEoverridecommandlockouts

\usepackage{cite}
\usepackage{amsmath,amssymb,amsfonts}
\usepackage{algorithmic}
\usepackage{graphicx}
\usepackage{textcomp}
\usepackage{xcolor}
\def\BibTeX{{\rm B\kern-.05em{\sc i\kern-.025em b}\kern-.08em
    T\kern-.1667em\lower.7ex\hbox{E}\kern-.125emX}}
\begin{document}

\title{D2-MLP: Dynamic Decomposed MLP Mixer for Medical Image Segmentation}

\author{
\IEEEauthorblockN{Jin Yang\IEEEauthorrefmark{1}, Xiaobing Yu\IEEEauthorrefmark{1}, Peijie Qiu\IEEEauthorrefmark{1}}
\\
\IEEEauthorblockA{\IEEEauthorrefmark{1}Mallinckrodt Institute of Radiology, Washington University School of Medicine,
St. Louis, USA \\
yang.jin@wustl.edu }}

\maketitle

\begin{abstract}
Convolutional neural networks are widely used in various segmentation tasks in medical images. However, they are challenged to learn global features adaptively due to the inherent locality of convolutional operations. In contrast, MLP Mixers are proposed as a backbone to learn global information across channels with low complexity. However, they cannot capture spatial features efficiently. Additionally, they lack effective mechanisms to fuse and mix features adaptively. To tackle these limitations, we propose a novel Dynamic Decomposed Mixer module. It is designed to employ novel Mixers to extract features and aggregate information across different spatial locations and channels. Additionally, it employs novel dynamic mixing mechanisms to model inter-dependencies between channel and spatial feature representations and to fuse them adaptively. Subsequently, we incorporate it into a U-shaped Transformer-based architecture to generate a novel network, termed the Dynamic Decomposed MLP Mixer. We evaluated it for medical image segmentation on two datasets, and it achieved superior segmentation performance than other state-of-the-art methods.

\end{abstract}

\begin{IEEEkeywords}
MLP Mixer, dynamic networks, medical image segmentation.
\end{IEEEkeywords}

\section{Introduction}
Segmentation of organs or lesions in medical images is crucial in supporting clinical workflows. However, manual segmentation is time-consuming and error-prone, thus motivating the development of automatic segmentation tools. Recently, Convolution Neural Networks (CNNs) have been widely used for automated medical image segmentation. Among CNN-based methods, U-Net and its variants are the most successful networks for medical image segmentation \cite{ronneberger2015u,cciccek20163d,zhou2018unet++,yang2023abdominal,yang2024dynamic}. However, their performance is limited by the inherent locality of convolutional operations due to the challenges in learning long-range semantic information.

To overcome the inherent limitations of CNNs, the Mixer is proposed to capture long-range information based on multi-layer perceptions (MLPs), achieving a competitive performance with CNNs \cite{tolstikhin2021mlp,cao2023strip}. Due to its performance and efficiency on computer vision tasks, the MLP Mixer is applied for segmentation tasks in medical images \cite{valanarasu2022unext,pan2023abdomen,pang20223d,liu2024rolling,zhou2023multibranch,shi2022polyp}. However, among these methods, some lack mechanisms to capture spatial features in basic MLP blocks, lowering the accuracy of dense predictions in medical images \cite{pang20223d,zhou2023multibranch,pan2023abdomen}. Others utilize some techniques, such as shifted MLPs or cycle MLPs, to learn spatial representations \cite{valanarasu2022unext,liu2024rolling}. However, the aggregation of information among different spatial locations and channels is insufficient. Additionally, they lack effective mechanisms to fuse spatial and channel features adaptively.

Dynamic mechanisms have been applied to adaptively capture features \cite{han2021dynamic}. Some methods employ dynamic mechanisms to adaptively aggregate features from multiple convolutional kernels based on their attention scores \cite{chen2020dynamic,li2019selective}. D-Net employs dynamic mechanisms to recalibrate and fuse features from different large kernels and levels \cite{yang2024d}. Dynamic Transformer employs a dynamic mechanism to fuse tokens from multiple windows \cite{ren2022beyond}. AgileFormer employs a dynamic mechanism to capture spatial features adaptively \cite{qiu2024agileformer}. However, few works apply dynamic mechanisms to adaptively aggregate and mix features in MLP Mixers for medical image segmentation.

To tackle these limitations, we propose a novel \textbf{Dynamic Decomposed Mixer} (DDM) module. The DDM module captures and aggregates features across different spatial locations and channels via two novel Mixers, including a \textbf{Spatially Decomposed Mixer} and a \textbf{Channel Mixer}. Specifically, our DDM module consists of three parallel paths. Two paths utilize the Spatially Decomposed Mixer to capture features and aggregate information along two different spatial dimensions, height and width, separately. It is achieved by decomposing input features into patches and rearranging them along height and width dimensions, separately. Then two MLPs are applied to capture information along with height and width, thus improving the extraction of spatial features across the whole spatial dimension. The third path employs a Channel Mixer to capture features along channels. Subsequently, the DDM module employs two dynamic mixing mechanisms, \textbf{Spatial-wise} and \textbf{Channel-wise Dynamic Mixing mechanisms} to model inter-dependencies between these channel and spatial features and to adaptively fuse them. Specifically, spatial features are extracted along two dimensions, the height and width, separately. Thus, to eliminate the isolation between these spatial features from two Spatially Decomposed Mixers, the Spatial-wise Dynamic Mixing mechanism is proposed to enhance their interactions and model inter-dependencies between spatial dimensions. The Channel-wise Dynamic Mixing mechanism is applied to adaptively fuse features from two Spatially Decomposed Mixers and the Channel Mixer.

We propose the \textbf{Dynamic Decomposed MLP Mixer} (D2-MLP) network for medical image segmentation by incorporating the DDM module into a hierarchical ViT-based encoder-decoder architecture. It can adopt behaviors of hierarchical Vision Transformers for learning hierarchical representations efficiently. We evaluated D2-MLP on two segmentation tasks, including Abdominal Multi-organ segmentation and Liver Tumor segmentation, and it achieved superior segmentation performance than state-of-the-art models. 

Our contributions have threefold: \textbf{(i)} We propose a novel Dynamic Decomposed Mixer module for learning representations. It is designed to capture features and aggregate information across different spatial locations and channels separately via the Spatially Decomposed Mixer and the Channel Mixer. Additionally, it employs novel Spatial-wise and Channel-wise Dynamic Mixing mechanisms to model inter-dependencies between spatial and channel features and to fuse them adaptively. \textbf{(ii)} We propose the Dynamic Decomposed MLP Mixer network by incorporating the Dynamic Decomposed Mixer module into a hierarchical ViT-based encoder-decoder for dense predictions. \textbf{(iii)} We evaluate the Dynamic Decomposed MLP Mixer network for medical image segmentation on two datasets. It achieved superior segmentation performance than other state-of-the-art methods.

\begin{figure*}[htbp]
\centerline{\includegraphics[width=0.95\textwidth]{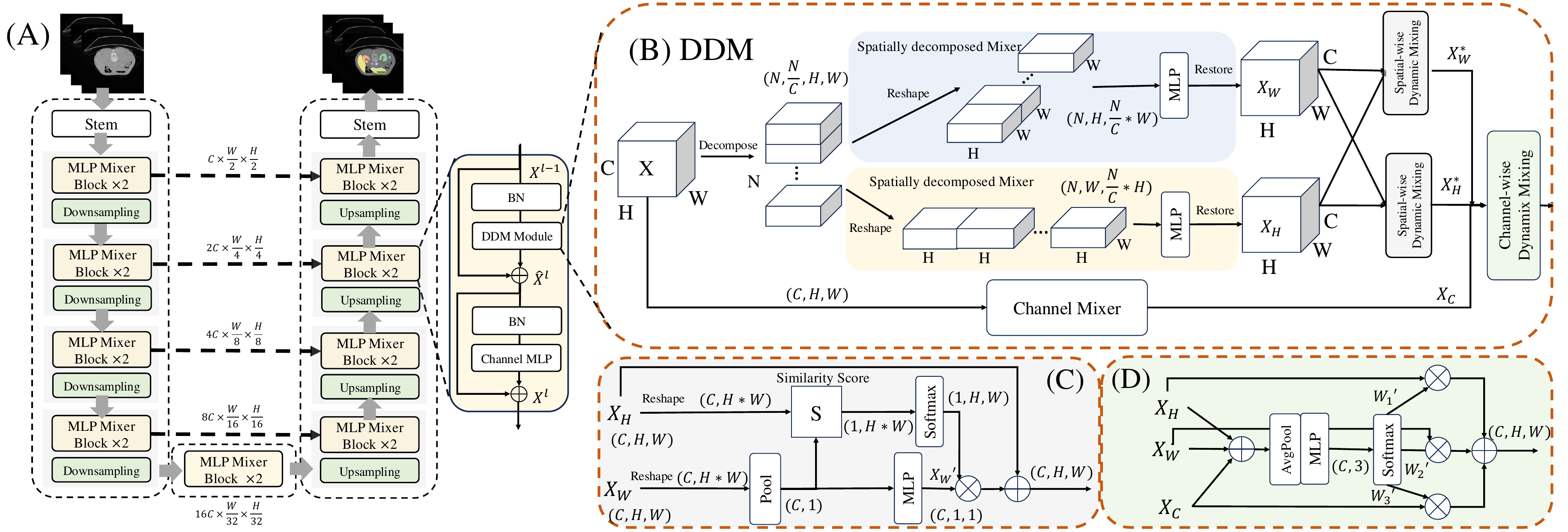}}
\caption{(A) D2-MLP is a 4-stage encoder-decoder architecture, and each MLP Mixer block consists of a DDM module and a channel MLP. (B) The DDM module decomposes the input feature $\boldsymbol{X}$ into $N$ patches. These patches are spatially reshaped and concatenated along height $H$ and width $W$ to features $\boldsymbol{X}_H$ and $\boldsymbol{X}_W$, separately. Subsequently, two MLPs are employed to aggregate information from $\boldsymbol{X}_H$ and $\boldsymbol{X}_W$ along two dimensions, separately. The interactions between $\boldsymbol{X}_H$ and $\boldsymbol{X}_W$ are improved in Spatial-wise Dynamic Mixing. A Channel Mixer is employed to aggregate information across channels from input features $\boldsymbol{X}$ as features $\boldsymbol{X}_C$. Lastly, features $\boldsymbol{X}_H^*$, $\boldsymbol{X}_W^*$, and $\boldsymbol{X}_C$ are adaptively fused in Channel-wise Dynamic Mixing. (C) Spatial-wise Dynamic Mixing. (D) Channel-wise Dynamic Mixing.}
\label{fig1}
\end{figure*}

\section{Methodology}
\subsection{Dynamic Decomposed Mixer Module}
The architecture of the DDM module is shown in Fig. \ref{fig1}.

\subsubsection{Spatially Decomposed Mixer} The first two paths utilize two Spatially Decomposed Mixers to aggregate spatial information from input features $\mathbf{X}\in\mathbb{R}^{C\times H\times W}$ ($C$: Channel; $H$: Height; $W$: Width) along two different spatial dimensions, $H$ and $W$, separately. Most MLP-based networks aggregate information in two dimensions simultaneously along channels, leading to inefficient token interaction and inflexible information aggregation. However, our Spatially Decomposed Mixer aggregates information along two spatial dimensions separately, resulting in more flexible feature interactions across different channels. 

Specifically, the Spatially Decomposed Mixer decomposes the input feature $\mathbf{X}$ into $N$ patches along channels, each with the dimension of $C'\times H\times W$ (where $C'=\frac{C}{N}$). The first path captures spatial features along the width $W$. These decomposed patches are spatially concatenated along the width $W$ and permuted to features $\mathbf{X}_W$ with the dimension of $(C'*W)\times N \times H$. Subsequently, an MLP, consisting of two linear layers, a $1\times3$ depthwise convolutional layer ($\mathrm{DWConv}$), and a GELU activation layer, is employed to capture features.
\begin{align}
    \nonumber
    \mathbf{X}_W &= \mathrm{Linear}(\mathbf{X}_W) \\
    \nonumber
    \mathbf{X}_W &= \mathrm{GELU}(\mathrm{DWConv}(\mathbf{X}_W)) \\
    \nonumber
    \mathbf{X}_W &= \mathrm{Linear}(\mathbf{X}_W)
\end{align}
In the second path, decomposed patches are spatially concatenated along the height $H$ and permuted to features $\mathbf{X}_H\in\mathbb{R}^{(C'*H)\times N \times W}$. Then the Spatially Decomposed Mixer is applied to capture features along the height $H$ by employing an MLP, consisting of two linear layers, a $1\times3$ depthwise convolutional layer, and a GELU activation.
\begin{align}
    \nonumber
    \mathbf{X}_H &= \mathrm{Linear}(\mathbf{X}_H) \\
    \nonumber
    \mathbf{X}_H &= \mathrm{GELU}(\mathrm{DWConv}(\mathbf{X}_H)) \\
    \nonumber
    \mathbf{X}_H &= \mathrm{Linear}(\mathbf{X}_H)
\end{align}
Then output features $\mathbf{X}_H$ and $\mathbf{X}_W$ are restored and transformed to the original dimension $C\times H\times W$.

\subsubsection{Channel Mixer} The third path employs a Channel Mixer to capture features $\boldsymbol{X}_C\in\mathbb{R}^{C\times H\times W}$ within each channel from the input $\mathbf{X}$. This Channel Mixer is an MLP consisting of two linear layers, a $3\times3$ depthwise convolutional layer, and a GELU activation layer.
\begin{align}
    \nonumber
    \mathbf{X}_C &= \mathrm{Linear}(\mathbf{X}) \\
    \nonumber
    \mathbf{X}_C &= \mathrm{GELU}(\mathrm{DWConv}(\mathbf{X}_C)) \\
    \nonumber
    \mathbf{X}_C &= \mathrm{Linear}(\mathbf{X}_C)
\end{align}

\subsubsection{Spatial-wise Dynamic Mixing} Two spatially decomposed features $\mathbf{X}_H$ and $\mathbf{X}_W$ are extracted along two dimensions in two isolated paths separately. Thus, we propose a Spatial-wise Dynamic Mixing mechanism to improve their interaction and model their correlations. First, we calculate the similarity score $\mathbf{S}\in\mathbb{R}^{1\times H\times W}$ to demonstrate the correlations between each feature in $\mathbf{X}_H$ and spatial-wise global information $\Bar{\mathbf{X}}_W$ ($\Bar{\mathbf{X}}_W$ is derived from $\mathbf{X}_W$ by calculating average values via pooling $\sum^n_{i=1}\frac{x_{(w,i)}}{n}$). Then this similarity score $\mathbf{S}$ is normalized and re-scaled by a Softmax function. To improve the generalizability of features $\mathbf{X}_H$, a tiny MLP is applied by cascading two linear layers with a GELU in between. Then the features are mixed based on their inter-dependencies dynamically as $\mathbf{X}_H^*\in\mathbb{R}^{C\times H\times W}$. A residual connection is also applied.
\begin{align}
    \nonumber
    \mathbf{S}&=\mathrm{Softmax}(\mathbf{X}_H\Bar{\mathbf{X}}_W) \\
    \nonumber
    \mathbf{X}'_W&=\mathrm{Linear}(\mathrm{GELU}(\mathrm{Linear}(\Bar{\mathbf{X}}_W))) \\
    \nonumber
    \mathbf{X}_H^* &= \mathbf{X}'_W \cdot \mathbf{S} + \mathbf{X}_H
\end{align}
We follow the same way to calculate $\mathbf{X}_W^*\in\mathbb{R}^{C\times H\times W}$ from $\mathbf{X}_W$ and spatially global information $\Bar{\mathbf{X}}_H$.
\begin{align}
    \nonumber
    \mathbf{S} &=\mathrm{Softmax}(\mathbf{X}_W\Bar{\mathbf{X}}_H) \\
    \nonumber
    \mathbf{X}'_H &=\mathrm{Linear}(\mathrm{GELU}(\mathrm{Linear}(\Bar{\mathbf{X}}_H))) \\
    \nonumber
    \mathbf{X}_W^* &= \mathbf{X}'_H \cdot \mathbf{S} + \mathbf{X}_W
\end{align}

\subsubsection{Channel-wise Dynamic Mixing} The Channel-wise Dynamic Mixing mechanism is applied to adaptively fuse spatial features $\mathbf{X}_H^*$ and $\mathbf{X}_W^*$ and channel features $\mathbf{X}_C$. Specifically, an adaptive average pooling is applied to calculate channel-wise significance scores $\mathbf{W}\in\mathbb{R}^{C\times1\times1}$. Then a tiny MLP network is employed to improve the descriptions of these scores by cascading two linear layers with a GELU activation in between. A Softmax function is utilized for normalization. Lastly, features are fused based on their significance scores as the output $\mathbf{X}_{out}\in\mathbb{R}^{C\times H\times W}$.
\begin{align}
    \nonumber
    &\mathbf{W} = \mathrm{AvgPool}(\mathbf{X}_H^*+\mathbf{X}_W^*+\mathbf{X}_C) \\
    \nonumber
    &\mathbf{W}' = \mathrm{Linear}(\mathrm{GELU}(\mathrm{Linear}(\mathbf{W}))) \\
    \nonumber
    &[\mathbf{W}'_1,\mathbf{W}'_2,\mathbf{W}'_3] =\mathrm{Softmax}(\mathbf{W}') \\
    \nonumber
    &\mathbf{X}_{out} = \mathbf{X}_H^* \cdot \mathbf{W}'_1 +\mathbf{X}_W^* \cdot \mathbf{W}'_2 + \mathbf{X}_C \cdot \mathbf{W}'_3
\end{align}

\subsection{MLP Mixer block}
The MLP Mixer block is the basic block for representation learning in segmentation networks. It is constructed by replacing the multi-head self-attention in a standard hierarchical ViT block with the DDM module (Fig. \ref{fig1}). The yielded Mixer block consists of a DDM module and a Channel MLP module. A Batch Normalization (BN) layer is applied before each DDM module and Channel MLP module. A residual connection is applied after each module. Thus, the MLP Mixer block in the $l$-th layer can be computed as
\begin{align}
    \nonumber
    \hat{\boldsymbol{X}}^l &= \textrm{DDM}(\textrm{BN}(\boldsymbol{X}^{l-1}))+\boldsymbol{X}^{l-1}, \\
    \nonumber
    \boldsymbol{X}^l &= \textrm{MLP}(\textrm{BN}(\hat{\boldsymbol{X}}^l))+\hat{\boldsymbol{X}}^l.
\end{align}

\subsection{Overall architecture}
The D2-MLP network is designed as a 4-stage U-shaped encoder-decoder architecture for learning hierarchical feature representations (Fig. \ref{fig1}). In the encoder, the stem employs a $7\times7$ convolutional layer with $2$ strides to partition the input images and project them to $C$ channels, thus generating features with the dimension of $C\times\frac{H}{2}\times\frac{W}{2}$. At each stage, two MLP Mixer blocks are stacked to perform representation learning, and a $2\times2$ convolutional layer with $2$ strides is employed to downscale the feature maps and increase the number of channels by a factor of 2. In the bottleneck, two consecutive MLP Mixer blocks are utilized. At each stage of the decoder, a $2\times2$ transposed convolutional layer with $2$ strides is employed to upscale feature maps and decrease the number of channels by a factor of 2. Subsequently, these upsampled features are concatenated with features from the same stage of the encoder via skip connections. Two consecutive MLP Mixer blocks are then utilized. In the stem of the decoder, a $2\times2$ transposed convolutional layer is employed. Lastly, a $1\times1$ convolutional layer is used to produce the dense segmentation predictions. The number of feature maps at each stage is $\{C, 2C, 4C, 8C, 16C\}=\{48, 96, 192, 384, 768\}$.

\begin{table*}
  \centering
    \caption{Comparison of segmentation performance between the D2-MLP and other SOTA approaches on the FLARE 2021 Multi-organ Segmentation dataset and MSD Liver Tumor Segmentation dataset. The best results are labeled by \textbf{Bold}. ($^*$: $p<0.01$ with Wilcoxon signed-rank test between D2-MLP and each SOTA method.)}
  \begin{tabular}{c|ccccccc|ccccc}
    \hline
    Methods & Dice$\uparrow$ & 95HD$\downarrow$ & MSD$\downarrow$ & Liver$\uparrow$ & Kidney$\uparrow$ & Spleen$\uparrow$ & Pancreas$\uparrow$ & Dice$\uparrow$ & 95HD$\downarrow$ & MSD$\downarrow$ & Liver$\uparrow$ & Tumor$\uparrow$ \\
    \hline
    Att U-Net & 91.61 & 4.71 & 1.04 & 97.81 & 96.07 & $\boldsymbol{97.15}$ & 75.43 & 73.10 & 16.96 & 5.33 & 94.71 & 51.49 \\
    nnU-Net & 91.54  & 4.74 & 1.05 & 97.83 & 96.06 & 97.14 & 75.14 & 72.50 & 19.49 & 7.90 & 94.49 & 50.51 \\
    DconnNet & 91.42 & 4.93 & 1.10 & 97.62 & 95.55 & 96.94 & 75.57 & 72.80 & 19.53 & 7.22 & 94.45 & 51.15  \\
    \hline
    Swin U-Net & 87.86 & 7.62 & 1.72 & 96.82 & 93.60 & 95.74 & 65.28 & 65.16 & 38.50 & 14.45 & 92.73 & 37.59 \\
    MISSFormer & 90.94 & 5.94 & 1.34 & 97.48 & 95.39 & 96.46 & 74.43 & 71.93 & 26.25 & 9.33 & 94.15 & 49.71 \\
    \hline
    UTNet & 89.20 & 6.68 & 1.47 & 97.32 & 94.94 & 96.52 & 68.03 & 71.16 & 21.26 & 9.50 & 93.85 & 48.47 \\
    UCTransNet & 91.84 & 4.93 & 1.09 & 97.81 & 96.10 & 96.81 & 76.63 & 72.04 & 21.14 & 7.59 & 93.81 & 50.27 \\
    HiFormer & 90.17 & 7.14 & 1.73 & 97.13 & 94.89 & 95.43 & 73.22 & 71.71 & 26.57 & 9.63 & 93.94 & 49.60 \\
    \hline
    UNeXt & 89.09 & 7.04 & 1.56 & 96.90 & 94.53 & 95.83 & 69.12 & 71.70 & 22.63 & 8.73 & 93.69 & 49.70  \\
    \hline
    D2-MLP & $\boldsymbol{92.53}^*$ & $\boldsymbol{3.99}^*$ & $\boldsymbol{1.02}^*$ & $\boldsymbol{98.21}$ & $\boldsymbol{96.39}$ & 96.52 & $\boldsymbol{79.00}$ & $\boldsymbol{75.73}^*$ & $\boldsymbol{15.25}^*$ & $\boldsymbol{4.93}^*$ & $\boldsymbol{95.37}$ & $\boldsymbol{56.10}$ \\
    \hline
  \end{tabular}
  \label{tab:1}
\end{table*}

\begin{figure*}[htbp]
\centerline{\includegraphics[width=0.9\textwidth]{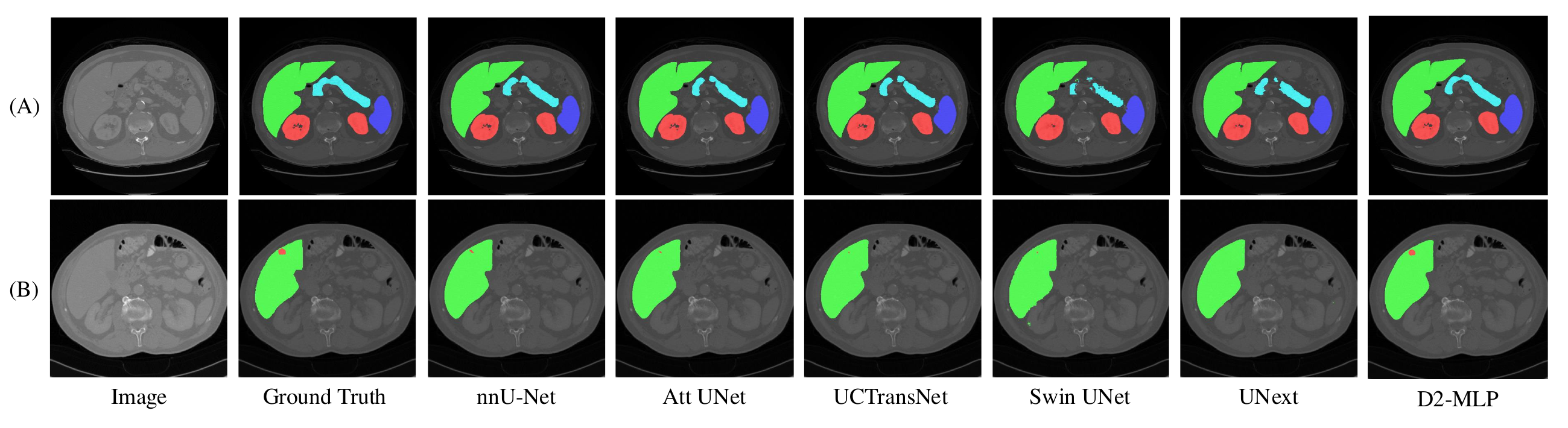}}
\caption{Qualitative comparison between D2-MLP and other methods in (A) the FLARE Multi-organ and (B) MSD Liver Tumor datasets.}
\label{fig2}
\end{figure*}

\section{Experiments}
\subsection{Datasets}
The first dataset is the FLARE 2021 multi-organ segmentation dataset which includes 361 multi-contrast CT images with voxel-wise manual annotations of four abdominal organs, including the liver, kidney, spleen, and pancreas \cite{ma2021abdomenct}. The second dataset is the Medical Segmentation Decathlon (MSD) Liver tumor segmentation dataset \cite{antonelli2022medical}. This dataset includes 131 Portal venous phase CT images with manual annotations of liver and liver tumors.

\subsection{Implementation details}
The D2-MLP is implemented using PyTorch. A combination of dice loss and cross-entropy loss was used as the loss function. The Stochastic Gradient Descent (SGD) was used as the optimizer. The initial learning rate was set to 0.001 and was decayed with a poly learning rate scheduler. The models were trained for 1000 epochs with deep supervision. The batch size was 14 and the input patch size was $512\times512$ in two datasets. 5-fold cross-validation was utilized to split each dataset and evaluate models.

\begin{table}
  \centering
  \caption{The comparison of performance among various designs of the D2-MLP method on the FLARE Multi-organ Segmentation dataset. The best Dice scores are labeled by \textbf{Bold}.}
  \begin{tabular}{c|c|cccc}
    \hline
    Models & Average & Liver & Kidney & Spleen & Pancreas \\
    \hline
    $N=2$  & 91.29 & 96.71 & 93.93 & 95.14 & 79.36 \\
    $N=4$  & $\boldsymbol{92.53}$ & $\boldsymbol{98.21}$ & $\boldsymbol{96.39}$ & $\boldsymbol{96.52}$ & 79.00 \\
    $N=8$  & 92.36 & 97.85 & 95.95 & 95.75 & $\boldsymbol{79.90}$ \\
    $N=16$ & 90.38 & 96.95 & 94.22 & 93.67 & 76.67 \\
    \hline
    Basic Mixer  & 88.36 & 94.64 & 93.02 & 92.58 & 73.22   \\
    DDM      & 92.53 & 98.21 & 96.39 & 96.52 & 79.00    \\
    \hline
  \end{tabular}
  \label{tab:2}
\end{table}

\subsection{Experimental results}
To evaluate the model performance, we employed the Dice coefficient (Dice), 95th Percentile Hausdorff Distance (95HD), and Mean Surface Distance (MSD) as evaluation metrics. To implement a thoughtful comparison, we compared D2-MLP with various 2D SOTA models, including CNN-based models (Attention U-Net \cite{schlemper2019attention}, nnU-Net \cite{isensee2021nnu}, and DconnNet \cite{yang2023directional}), ViT-based models (Swin U-Net \cite{cao2022swin} and MISSFormer \cite{huang2022missformer}), hybrid ViT-CNN models (UTNet \cite{gao2021utnet}, UCTransNet \cite{wang2022uctransnet}, and HiFormer \cite{heidari2023hiformer}), and a MLP-based model (UNeXt \cite{valanarasu2022unext}). Table \ref{tab:1} shows that the D2-MLP network achieved superior overall performance over other SOTA methods on both two segmentation tasks. The D2-MLP model showed significant improvement across almost all organ-specific segmentation tasks. The qualitative comparison shows that D2-MLP achieved better results than other SOTA methods (Fig. \ref{fig2}). 

\subsection{Ablation study on Dynamic Decomposed Mixer module}
\subsubsection{The impact of patch number} We conducted an ablation study to investigate the impact of the patch number $N$ on model performance. Table \ref{tab:2} shows that the D2-MLP achieved the best segmentation performance when the patch number is 4 ($N=4$). It achieved the second-best performance when the patch number is 8 ($N=8$).

\subsubsection{The effectiveness of DDM module}
In this study, we evaluated the effectiveness of the DDM module on medical image segmentation by replacing it with a basic Channel Mixer module in D2-MLP. Table \ref{tab:2} demonstrates that D2-MLP with the DDM module achieved a much higher Dice score than that with a basic Mixer module, showing its effectiveness on medical image segmentation.

\section{Conclusions}
We propose a Dynamic Decomposed MLP Mixer network for medical image segmentation. This network employs a Dynamic Decomposed Mixer module to learn spatial and channel features and aggregate them adaptively. The experimental results demonstrate the superior performance of our segmentation model over other SOTA methods and the benefits of the Dynamic Decomposed Mixer module on segmentation.

\bibliographystyle{ieeetr}
\bibliography{MLP}

\end{document}